\documentclass[aps,prd,twocolumn,showpacs,preprintnumbers]{revtex4}

\usepackage{amsmath,amssymb}
\usepackage{graphicx}

\newcommand{\lp}{\ell_{\mathrm P}}
\newcommand{\be}{\begin{equation}}
\newcommand{\ee}{\end{equation}}
\newcommand{\bq}{\begin{eqnarray}}
\newcommand{\eq}{\end{eqnarray}}
\newcommand{\snf}{\sin(\bar \mu c)}
\newcommand{\snfs}{\sin^2(\bar \mu c)}
\newcommand{\csf}{\cos(\bar \mu c)}

\newcommand{\heff}{{\cal H}_{\mathrm{eff}}}

\newcommand{\hcl}{{\cal H}_{\mathrm{cl}}}

\newcommand{\hm}{{\cal H}_{\mathrm{M}}}

\newcommand{\rcr}{\rho_{\mathrm{crit}}}

\newcommand{\lra}{\longrightarrow}

\newcommand{\f}{\frac}
\newcommand{\n}{\nonumber}
\newcommand{\xc}{x_{\mathrm{cl}}}
\newcommand{\yc}{y_{\mathrm{cl}}}
\newcommand{\lc}{\Omega_{\mathrm{cl}}}

\begin{document}

\preprint{IGPG-06/2-3}

\title{Loop cosmological dynamics and dualities with 
Randall-Sundrum braneworlds}
\author{Parampreet Singh\footnote{e-mail address: {\tt singh@gravity.psu.edu}}}
\affiliation{Institute for Gravitational Physics and Geometry, Pennsylvania State University,
University Park, PA 16802, USA}

\begin{abstract}
The discrete quantum geometric effects play an important role in
dynamical evolution in the  loop quantum cosmology. These
effects which are significant  at the high energies lead to the quadratic energy density   
modifications to the Friedmann equation, as in the
Randall-Sundrum braneworld scenarios but with a negative sign.  We investigate the scalar field dynamics in this scenario and 
show the existence of a phase of super-inflation independent of the inverse scale factor modifications as found earlier.
In this regime the scalar field mimics the dynamics of a  phantom field and vice versa.
We also find various
symmetries between the expanding phase, the contracting phase and the
phantom phase in the loop quantum cosmology. We then 
construct the scaling 
solutions in the loop quantum cosmology and show their dual
relationship with those of the Randall-Sundrum cosmology.

\end{abstract}

\pacs{98.80.Cq,04.60.Pp}

\maketitle

\section{Introduction}

The  standard  model  of  cosmology  has successfully  provided  us  a
consistent  picture  of  the  evolution  of our  Universe  in  various
epochs. However,  it is  expected that when  the limit of  validity of
general relativity is  reached, in the regime of  very high curvature,
the standard Friedmann dynamics shall be modified. It thus has been an
attractive idea  to seek modifications  to the Friedmann  equations at
the high  energy scales. A  theory of quantum gravity  shall naturally
provide  us  these modifications.   Motivated  by  the string  theory,
modifications to  the Friedmann  dynamics in the  Randall-Sundrum (RS)
braneworld   scenarios  \cite{rs}   have   been  extensively   studied
\cite{roy}.  Randall-Sundrum  braneworld scenarios  are  based on  the
Horova-Witten model  \cite{hw} where  a $3+1$ dimensional  universe is
obtained  after   compactification  of  a   6  dimensional  Calabi-Yau
manifold.  The bulk spacetime in Randall-Sundrum model is 5 dimensional 
anti-de Sitter with the extra dimension being spacelike and large.  
The anti-de  Sitter  bulk in  the Randall-Sundrum  scenario
leads to  the localization  of gravity on  the brane and  the modified
Friedmann  equation on the  brane is of  the form  $H^2 \propto  \rho(1 +
\rho/2\sigma)$  where $H$  is the  Hubble rate,  $\rho$ is  the energy
density of matter and $\sigma$ is the brane tension. The $\rho^2$ modification 
in the effective Friedmann equation is directly an effect of the existence of 
a large extra dimension in this model, as it arises by using the Israel junction conditions
on the bulk-brane system.

Modifications to the Friedmann dynamics appear also in loop quantum
cosmology (LQC) \cite{martin} which is the symmetry reduced
quantization of homogeneous and isotropic spacetimes based on the loop
quantum gravity (LQG) \cite{lqg_review}.  LQG which is one of the
background independent and non-perturbative candidate theories for
quantizing Einsteinian gravity in four dimensions predicts that at quantum level, classical spacetime
continuum is replaced by a discrete quantum geometry and operators corresponding to 
length of a curve, area of a surface and volume of an enclosed region have discrete
eigenvalues.  The geometrical
operators in LQC, for example the scale factor and the inverse scale
factor, also have a discrete spectrum and the underlying dynamics in
loop quantum cosmology is governed by a non-singular discrete quantum difference
equation \cite{martin}. However recent investigations, pertaining to the study of
evolution of the semi-classical states have shown that the discrete
quantum dynamics can be very well approximated by an effective
modified Friedmann dynamics till scales very close to the Planck scale
\cite{time,eff_diff,aps}. The modifications to the Friedmann dynamics
due to loop quantum effects are of two types. First is based on the
modification to the behavior of inverse scale factor below a critical
scale factor $a_*$ determined by a half integer parameter $j$. This
parameter arises because inverse scale factor operator is computed by
tracing over SU(2) holonomies in an irreducible spin $j$
representation. It turns out that the eigenvalues of the inverse scale
factor operator become proportional to the positive powers of scale
factor for $a \lesssim a_*$.  This change regulates the divergence of energy
density for arbitrary matter \cite{freq} and changes the classical frictional term to anti-frictional in the 
Klein-Gordon equation for the scalar field in an expanding Universe, thus leading to a phase of super-inflation \cite{superinflation}.
Various interesting applications have been found, for example
resolution of big bang singularity by loop quantum dynamics
\cite{bigbang}, avoidance of many  singularities in cosmological \cite{cosmosing} and
gravitational collapse scenarios \cite{gravsing} by inverse scale
factor modifications, increasing the
viability of the onset of inflation \cite{superinflation,cmb,inflation}, non-singular
cyclic models \cite{cyclic}, natural trans-Planckian modifications to
the frequency dispersion relation \cite{freq} etc.

The second type of modification essentially encodes the discrete
quantum geometric nature of spacetime, as predicted by the loop
quantum gravity, in the Friedmann dynamics. As we would discuss in the next section, this modification arises
because the loops on which holonomies are computed have a
non-vanishing minimum area given by the eigenvalues of the area
operator in LQG \cite{eff_diff,time,aps,aps2,wkb}.  It leads to a $\rho^2$ modification of the Friedmann
equation of the form, $H^2 \propto \rho(1 - \rho/\rcr)$ where $\rcr^{-1} =
\alpha \kappa \gamma^2 \lp^2/3$ and $\rho$ in general has
modifications due to inverse scale factor for $a < a_*$. Here $\kappa
= 8 \pi G$ with $G$ being the four dimensional gravitational constant,
$\lp = \sqrt{G}$ is the Planck length \cite{formula}, $\gamma \approx
0.2375$  is the dimensionless Barbero-Immirzi parameter whose value is set by the
black hole thermodynamics in  LQG \cite{bek_hawking} and $\alpha$ is a
constant of the order unity determined by the eigenvalues of  the area
operator. The form of the modification in the Friedmann equation leads
to a non-singular bouncing cosmology \cite{aps}.

For small values of $j$ parameter or for scale factors $a > a_*$, the modifications to the effective Friedmann dynamics 
due to change in the behavior of the inverse scale factor can be neglected and only those originating from discreteness effects 
are important. Working in this setting we 
would consider the evolution for a Universe composed of a massive
scalar field $\phi$ with a conjugate momentum $\Pi_\phi$ and matter
component with a constant equation of  state $w = p_w/\rho_w$. Our
consideration of matter in LQC would  be on the similar
phenomenological lines as in Ref. \cite{freq}. 

We would  aim to investigate various features pertaining to the effective dynamics in this paper. After deriving the necessary effective equations in Sec. II, we would analyze in detail the 
scalar field dynamics in Sec. III and show that LQC generically leads to a phase of super-inflation, independent of the inverse scale factor 
modifications, when $\rho > \rcr/2$.   Most of the phenomenological applications in LQC are based on the existence of this phase originating from the 
change in behavior of inverse scale factor for $a < a_*$. However, if value of $j$ is chosen to be small then most of these effects 
become weak. Our result about the existence of this phase which originates due to discrete quantum effects establishes the robustness 
of these applications. By considering the dynamical evolution of a massive scalar field and a phantom field in LQC, we would show a peculiar 
relation between them. In the regime $\rho > \rcr/2$, a massive scalar field behaves as a 
phantom field in the standard cosmology, and a phantom field mimics dynamics of  an ordinary scalar field. 
We would then  investigate various symmetries between the expanding branch, the contracting branch and the dynamics
of a phantom scalar field  and show that as in  the
standard Friedmann cosmology \cite{phantomduality,triality} and the Randall-Sundrum scenario \cite{triality}, they
have a dual relationship with each other in  the effective theory of
LQC. 

One of the most important questions pertinent for any dynamical evolution is its stability. For that we would 
derive scaling solutions in LQC  which are useful to study the stability
properties of cosmological models  \cite{clw,mizuno,scaling,sc1}.  As
has been further established in Ref. \cite{scaling}, these scaling
solutions may provide useful links between  distinct cosmological
scenarios. We explore this avenue in Sec. IV and find that the scaling solutions
in LQC have a dual relationship with those in the Randall-Sundrum
scenario. We shall note that the scaling solutions in LQC have been
constructed earlier \cite{lidsey,clm} and  correspondences between
string inspired scenarios and loop inspired cosmologies have been
found \cite{clm}, though in the case when only inverse scale factor
modifications to the  effective dynamics are important and the
discrete quantum gravity effects play no role (which may happen if
$\rho \ll \rcr$ in all phases of the cosmological evolution). In this
sense the scaling solutions found here are complimentary to those
in Refs. \cite{lidsey,clm}.  

In Sec V, we summarize the results obtained in this paper and discuss the implications for the duality symmetry between scaling solutions of 
LQC and Randall-Sundrum scenario. We would discuss the way this relationship can be used for various useful applications and to extract 
physical predictions, for example the perturbation spectrum in LQC. Just with its use as a mathematical device this duality can be used to investigate the 
detailed properties of solutions in LQC given those in Randall-Sundrum scenario. However we would also discuss if it points out to a deeper 
relationship between two frameworks and its ramifications.

\section{Effective dynamics  in loop quantum cosmology}
LQG is a quantization of gravity based on Ashtekar-Barbero connection variables
with the gravitational phase space spanned by SU(2) connection $A^i_a$ and the triad 
$E^a_i$ on a 3-manifold M (labels $a$ and  $i$ denote space and internal indices respectively). 
In LQC, on imposition of symmetries of the framework, 
 the non-trivial information about the classical phase space gets encoded in variables $c$ related to $A^i_a$ 
and $p$ related to the triad $E^a_i$ which can have two orientations. On classical solutions (of general relativity) $c$ is given by $c = k + \gamma \dot a$, $k$ being curvature index which we would take to be zero in this work. The triad $p$ is   
related to the scale  factor $a$ of the homogeneous and isotropic metric via $|p| = a^2$, 
where the modulus  arises due to two possible orientations of $p$ and in this work we choose 
the positive one without any loss of generality. 
In LQG, the connection does not have a corresponding quantum operator hence it is more useful to 
work with holonomies defined over a loop. 
The holonomy
over an edge $\epsilon$ of a loop is defined as $h_i := \cos (\mu c/2) + 2 \tau_i  \sin (\mu c/2)$ where  
$\tau_i$ are related to the Pauli spin matrices as $\tau_i = - i \sigma_i/2$, 
and dimensionless $\mu$ is related to the physical length ${\lambda}_\epsilon$ of the edge $\epsilon$ as 
${\lambda}_\epsilon = \mu \, |p|^{1/2}$.

Given this classical phase space structure, we then perform quantization on the lines of LQG by promoting 
holonomies and triads to quantum operators \cite{Bohr}. It turns out that the triad operator $\hat p$ has 
 a discrete eigenvalue spectrum with eigenvalues $\mu$,
\be
\hat p \, |\mu \rangle = \f{4 \pi \mu \gamma \lp^2}{3} \, |\mu \rangle 
\ee
including the eigenvalue zero. Thus the naive inverse of the triad operator is not densely defined. 
The eigenvalues of the inverse triad operator are important as they give us information about the way curvature which is proportional to the inverse powers of the scale factor (or equivalently the triad) 
would behave in LQC. In order to evaluate it we use a classical identity of the Ashtekar-Barbero phase space 
\be \label{overpeq}
\f{1}{\sqrt{|p|}} = \f{1}{2 \pi G \gamma} \, {\mathrm{tr}} \left(\sum_i \, \tau^i h_i \{ h_i^{-1}, V^{1/3} \} \right) ~.
\ee
Here $V$ denotes volume related to $p$ as  $V = |p|^{3/2}$. 
The operator  $\widehat{(\sqrt{|p|})^{-1}}$ commutes with the operator $\hat p$ and has eigenstates  $|\mu \rangle$. 
It can be shown that its eigenvalue spectrum is bounded on the entire Hilbert space \cite{Bohr}. This implies that the curvature in LQC remains finite and does not 
blow up, even for the state $|\mu = 0 \rangle$ which corresponds to $a = 0$ (the classical big bang).

The Hamiltonian constraint operator is made of the  gravitational and the matter part, $\widehat{\cal H}
= \widehat{\cal H}_{\mathrm{G}} + \widehat{\cal H}_{\mathrm{M}}$. The classical gravitational constraint 
consists of inverse powers of the triad and curvature which are expressed in terms holonomies and their 
Poisson brackets with positive powers of $V$ (as in eq.(\ref{overpeq})).
The quantum operator for the gravitational part of the classical Hamiltonian constraint 
 is given by \cite{Bohr} 
\bq
\widehat{\cal H}_{\mathrm{G}} &=& \n \f{i}{4 \pi \kappa \lp^2 \gamma^3 \bar \mu^3} \, \sum_{ijk} \epsilon^{ijk} {\mathrm{ tr}} \left(\hat h_i \hat h_j \hat h_i^{-1} \hat h_j^{-1} \hat h_k [\hat h_k^{-1}, \hat V] \right) \\
&=& \f{6 i}{\pi \kappa \lp^2 \gamma^3 \bar \mu^3} \, \sin^2\left(\f{\bar \mu c}{2}\right) \cos^2\left( \f{\bar \mu c}{2}\right)\\
&& \n \times \left(\sin\left(\f{\bar \mu c}{2}\right) \hat V \cos \left(\f{\bar \mu c}{2}\right) - \cos\left(\f{\bar \mu c}{2}\right)  \hat V \sin\left(\f{\bar \mu c}{2}\right) \right)
\eq
where we have used the definition of the holonomies. These holonomies are computed over square loops with physical area given by the 
 minimum eigenvalue of the area operator in LQG which is  $\alpha \lp^2$ where $\alpha$ is of the order
unity. We have denoted the physical length of an edge of such a loop by $\bar \mu |p|^{1/2}$ with the area of the loop ${\cal A} = \bar \mu^2 |p| = \bar \mu^2 a^2$, then its equality with minimum eigenvalue of area operator in LQG yields
\be\label{mubar}
 \bar \mu^2 a^2 = \alpha \lp^2 ~.
\ee
 Thus quantum geometry acts like a regulator for the size of the loops over which holonomies are evaluated and
brings in elements of quantum discreteness inherited from LQG.

Using the relation $V = |p|^{3/2}$, we can also define the volume and the inverse volume operator  which 
lead to a discrete eigenvalues of $\hat V$ given by $V_{\mu} \equiv (4 \pi  \mu \gamma \lp^2/3)^{3/2}$  and of $\widehat{V^{-1}}$ given by \cite{superinflation}
\be \label{dj}
d_j(\mu) = \Bigg[\f{4}{3 \pi \lp^2 \gamma \bar \mu j(j + 1)(2 j + 1)} \sum_{n = - j}^{j} \, n V_{\mu + 2 n \bar \mu}^{1/2} \Bigg]^6
\ee
which is bounded and implies that physical densities remain finite in LQC.
Here $\bar \mu$ is given by
eq.(\ref{mubar}).
The parameter $j$ arises due to tracing over
SU(2) holonomies in an irreducible spin $j$ representation. 
In particular it determines a critical scale factor
 $a_* = \sqrt{8\pi \gamma j \bar\mu/3} \, \lp$ below which the eigenvalues of inverse scale factor operator 
become 
proportional to the positive powers of scale factor \cite{martin}. However, for $a \gtrsim a_*$, the eigenvalues $d_{j}$ quickly approximate the classical behavior i.e. $d_{j} \approx a^{-3}$.

The physical states are given by 
$|\psi\rangle = \sum_\mu \psi(\mu,\phi) |\mu\rangle$, where $\phi$ denotes matter degrees of freedom which constitute
the matter Hamiltonian operator ${\widehat{\cal H}_{\mathrm{M}}}$. The action of these states on $\widehat{\cal H}$
yields the quantum evolution given by the following discrete difference equation
\begin{widetext}
\begin{eqnarray}\label{seq}
	\frac{3}{4 \kappa \gamma^2 \bar\mu} \Big[
	C(\mu+4 \bar\mu) \,\psi(\mu + 4 \bar\mu, \phi) - 
	2 C(\mu) \, \psi(\mu, \phi) + C(\mu - 4 \bar \mu) \, \psi(\mu - 4 \bar \mu, \phi) \Big] + \hat{\cal H}_{\mathrm{M}} \psi(\mu, \phi) = 0 ~
\end{eqnarray}
\end{widetext}
with
\be
C(\mu) = \frac{1}{4 \pi \gamma \bar \mu \lp^2}(V_{\mu+\bar\mu}-V_{\mu-\bar\mu}) ~.
\ee
The quantum evolution determined by the above equation has been shown to be non-singular at the big bang \cite{bigbang,Bohr,aps}.
At the classical level, $\hm$ in general  
would contain inverse scale factors which would blow up when $a \lra
0$ and the evolution would break down. This is cured in LQC due to modification to the behavior 
of the eigenvalues of the inverse volume operator  by eq.(\ref{dj}) which remain bounded through the whole evolution.

In order to compare the discrete dynamics resulting from
the difference equation and the classical theory, one can construct 
semi-classical states, study their discrete quantum evolution and
compute the expectation values of the observables \cite{time,eff_diff,aps,aps2}. These
investigations show that for scale factors greater than of the order Planck length  we can consider the emergence of a continuous spacetime picture 
with the 
dynamics governed by  the following effective Hamiltonian constraint which 
approximates very well the evolution via the difference equation
\cite{eff_diff,aps2,eff_footnote}
\be\label{heff1}
\heff = - \f{3}{\gamma^2 \kappa \bar \mu^2} s_j \, \sin^2(\bar \mu c) + \hm
\ee
where $s_j$ is given by \cite{kevin_ham}
\be
s_j  = - \f{3}{4 \pi \lp^2 \gamma \bar \mu j (j + 1) (2 j + 1)} \sum_{n = -j}^{j} \, n V_{\mu - 2 n \bar \mu}
\ee
with a behavior similar to that of $d_j(\mu)$. For $a > a_*$ it is very well approximated by $s_j \approx a$.  Effects due to 
quantum geometric regulator for the minimum area of loops over which holonomies are computed are manifest in the 
gravitational part of $\heff$. The matter Hamiltonian  $\hm$  in general includes the modified behavior of the 
inverse scale factor (eq.(\ref{dj})).

We can compare the effective Hamiltonian obtained in LQC with the classical Hamiltonian constraint,
\be \label{hclass}
\hcl =  - \f{3}{\kappa} \dot a^2 \, a +  \hm = - \f{3}{\gamma^2 \kappa} c^2 \, p^{1/2} +  \hm ~.
\ee 
where we have used the relation $c =  \gamma \dot a$.
In the limit when $a \gg a_*$ and $\bar \mu c \ll 1$ it is easily seen that eq.(\ref{heff1}) reduces to eq.(\ref{hclass}).
We should note  that the modifications in the effective Hamiltonian  due to the
change in behavior of eigenvalues of the inverse  scale factor
operator are of significance if $a \lesssim a_*$, where as those due
to discrete quantum effects become important whenever $\bar \mu c$ is large. Since 
$\bar\mu \propto a^{-1}$ and $c \propto \dot a$, the discrete quantum effects become 
significant at large values of $H$ or $\rho^{1/2}$, strictly speaking when $\rho$ becomes of the 
order of $\rcr$ in eq.(\ref{mod}). The domain in which inverse scale factor modifications can be 
important is determined solely by the parameter $j$  which determines $a_*$, whereas the domain in which 
discrete quantum effects are important depends on the value of energy density. For a general choice of matter configuration it is  possible that inverse scale factor modifications 
and discrete quantum effects are significant in distinct domains. To see this let us consider an example of 
energy density $(\rho_\phi)$ sourced by a massless scalar field. Then $\rho_\phi \sim \rcr$ implies a 
critical scale factor $a_{\mathrm{crit}}$ near which discrete quantum geometric effects become significant.
This critical scale factor can be easily computed to be $a_{\mathrm{crit}} = (8 \pi \alpha \gamma^2/6)^{1/6} \Pi_\phi^{1/3} \lp $ where $\Pi_\phi$ is the scalar field momenta in Planck units. Further we can compute 
$a_*$ which on using eq.(\ref{mubar})  turns out to be $a_* = (8 \pi \alpha^{1/2}\gamma/3)^{1/3} j^{1/3} \lp$. 
Comparing the two scales we find that $a_{\mathrm{crit}} > a_*$ for $\Pi_\phi > 4 j$. Thus for these values of 
$\Pi_\phi$ we have a regime where discrete quantum effects are important and the modifications due to inverse 
scale factor can be ignored in the dynamics. However we should also note that even for values of $\Pi_\phi$ 
such that $a_{\mathrm{crit}} < a_*$, numerical investigations indicate that  the discrete quantum effects dominate those due to inverse scale factor, unless we choose an initial matter configuration such that 
$\bar \mu c \ll 1$ through out the evolution \cite{eff_diff}. The effects due to inverse scale factor 
modifications are further weakened when compared to discrete quantum effects if we take into account 
theoretical arguments which favor  a small value of $j$ parameter \cite{alex}. 
For these reasons in this work we would focus
exclusively on the modifications due to discrete quantum geometry
effects present in the first term of eq.(\ref{heff1}). This choice can be made 
by considering the matter configuration for a given value of $j$ such that $a_{\mathrm{crit}} > a_*$.
Then  modification due to inverse scale factor
can be neglected in (\ref{heff1}), especially the matter Hamiltonian $\hm$ and the corresponding expressions for energy density and pressure remain
same as classically \cite{freq}.

In general the matter Hamiltonian, $\hm$, would be composed of a 
massive scalar field $\phi$ with a conjugate momentum $\Pi_\phi$, 
energy density $\rho_\phi$ and a matter component with a constant 
equation of state $w = p_w/\rho_w$, and thus the total energy density $\rho = \rho_\phi + \rho_w$.
 For our case of interest, $a > a_*$,
 the matter Hamiltonian in the modified dynamics is 
given by \cite{freq}
\be
\hm = \frac{1}{2} \frac{\Pi_\phi^2}{p^{3/2}} + p^{3/2} \, V(\phi) + \rho_w \, p^{3/2} ~.
\ee
Also for this case the energy density and pressure for the scalar field 
are equal to their classical values \cite{freq}, i.e. 
\be \label{edp}
\rho_\phi = \frac{1}{2} \, \dot \phi^2 + V(\phi), ~~~ p_\phi = \frac{1}{2} \, \dot \phi^2 - V(\phi) ~
\ee
where we have used the Hamilton's equation $\dot \phi = \Pi_\phi/p^{3/2}$. It is then straightforward to find the Klein-Gordon equation using 
the Hamilton's equations for $\dot \phi$ and $\dot \Pi_\phi$, which turns out to be of the same form as the one 
classically and satisfies the stress-energy conservation law, $\dot \rho_\phi + 3 (\dot a/a) (\rho_\phi + p_\phi) = 0$.

The effective Hamiltonian which is now given by 
\be
\heff = - \f{3}{\gamma^2 \kappa \bar \mu^2}  \,a \sin^2(\bar \mu c) + \hm
\ee
leads via the Hamliton's equation of $\dot p$,
\be
\dot p = \{p,\heff\} = - \f{\gamma \kappa}{3} \f{\partial}{\partial c}\heff
\ee
the rate of change of the scale factor
\be
\dot a = \f{1}{\gamma \bar \mu}\snf \, \csf ~. \label{dota} 
\ee
Further, the vanishing of the Hamiltonian constraint implies
\be
\snfs = \frac{\kappa \gamma^2 \bar \mu^2}{3  \, a} \, \hm
\ee
which on using eq.(\ref{dota}) yields 
\be
H^2 = \frac{\kappa}{3} \, \rho \left( 1 - \f{\rho}{\rcr} \right), \, \, \rcr^{-1} = \alpha \kappa \gamma^2 \lp^2/3 \label{mod}
\ee 
where we have used eq.(\ref{mubar}). 

The modifications originating due to discrete quantum effects to the effective Friedmann equation in LQC are thus of the type 
where a $\rho^2$ term becomes important in the regime
of high energies. This feature holds even  at scales less than $a_*$ with $\rho$ now including modifications due to the peculiar behavior of eigenvalues of the inverse scale factor operator. When energy density becomes small compared to $\rcr$, the modification $\rho/\rcr$ becomes very small and the 
effective Friedmann equation reduces to the classical Friedmann equation.

Comparing with the string inspired 
Randall-Sundrum scenario, the sign of the correction term is negative which leads to a non-singular bouncing cosmology
\cite{aps}. Interestingly, similar Friedmann equation  arises in a braneworld scenario which is a  modification of the Randall-Sundrum model in the sense that the extra dimension is considered to be time-like \cite{shtanov}. As we would see the correspondence between the effective dynamics in LQC and the Randall-Sundrum braneworlds is much deeper, with the dual relationships between the scaling solutions of  
both of the scenarios.

\section{Scalar field dynamics}

Let us consider the dynamics in the effective theory for a
Universe with only a massive scalar field contributing as the matter component.
We further work with a generalization that we would 
allow the scalar field to be of phantom type which has
a negative kinetic energy \cite{phantom}. The energy density  and the
pressure in eq.(\ref{edp}) then generalize to 
\be
\rho_\phi = n \f{\dot \phi^2}{2} + V(\phi), \,\,p_\phi= n \f{\dot
  \phi^2}{2} - V(\phi)
\ee
where $n = \pm 1$ for the standard scalar field and the phantom field
respectively. The Klein-Gordon equation can then be written as 
\be \label{kgeq1}
\dot \rho_\phi = - 3 H \,(\rho_\phi + p_\phi) = - 3 n H \dot \phi^2 ~.
\ee

Using eq.(\ref{mod}), we can find the rate of change of the Hubble rate 
\be
\dot H = - \f{\kappa}{2} \, n \dot\phi^2 \left(1 - 2
\f{\rho}{\rcr} \right)
\ee
and 
\be
\f{\ddot a}{a} =  \f{\kappa}{3}\, \Bigg[\rho\left(1-\f{\rho}{\rcr} \right) -
  \f{3 n}{2} \dot\phi^2 \left(1 - 2
\f{\rho}{\rcr} \right) \Bigg] ~.
\ee
It is easy to then see from these equations dynamical features of the 
effective theory.
 In the case when $\rho \ll\rcr$, the dynamical evolution is classical. 
For the case of a standard scalar field, $\dot H <0$ and the sign of
$\ddot a$ depends on whether or not the potential term dominates over 
the kinetic term.
Whereas for the phantom field, $\dot H > 0$ and $\ddot a$ is positive
leading to a phase of super-inflation.

The effect of $\rcr$ term on the scalar field dynamics is very
peculiar, especially when $\rcr/2 < \rho < \rcr$. In this case,
for the standard scalar field $\dot H > 0$ and $\ddot a > 0$ 
leading to a phase of super-inflation, irrespective of the
choice of the potential.  However, for the
phantom field $\dot H < 0$ and the sign of the $\dot a$ term would now
depend on the ratio of potential to the kinetic energy, as for the
standard scalar field in standard Friedmann cosmology.
The behavior of the massive scalar field in the regime when loop
quantum modifications are very significant ($\rcr/2 < \rho < \rcr$) 
mimics that of a phantom field and the phantom field in this regime 
behaves like an ordinary scalar field. 

We note that for $a < a_*$, existence of a super-inflationary phase in 
LQC due to modifications pertaining to the inverse scale factor
behavior is known \cite{superinflation} and has been investigated in 
detail \cite{inflation,cmb}. It has been established that this phase can yield generic 
conditions for chaotic inflation to start even when the inflaton is initially 
at the bottom of the potential. Further, it also leads to distinct
signatures in the cosmic microwave background \cite{cmb}. However,
here we have shown that this phase generically exists in LQC even for 
$a > a_*$ due to discrete quantum geometric effects. This would
further increase the range of initial parameters for the onset
of inflation. For $a < a_*$, it would lead to an additional 
super-inflation and thus increase the number of e-foldings in the
loop modified phase.

In classical as well as the Randall-Sundrum braneworld cosmology, it has been shown that 
different phases of the scalar field dynamics are related by some symmetry transformations \cite{triality}.
We would now investigate this issue for the loop cosmology. For that we define 
a new variable
\be \label{hphi1}
h(\phi) = \f{\rho}{\rcr - \rho}
\ee
such that the Friedmann equation (\ref{mod}) can be written as
\be \label{cllm_hub}
a(\phi)_{,\phi} \, h(\phi)_{,\phi} = - n \kappa \, a(\phi) \,  h(\phi)
\ee
with $a(\phi)$ determined as
\be \label{aphi1}
a(\phi) = \exp\left(-\kappa \int d \phi \, h(\phi) h_{,\phi}^{-1} \right) ~.
\ee

An important property of eq.(\ref{cllm_hub}) is that two distinct cosmological models/phases 
represented by  $(a_1(\phi), h_1(\phi), V_1(\phi))$ and  $(a_2(\phi), h_2(\phi), V_2(\phi))$, leave 
eq.(\ref{cllm_hub}) invariant if
\be \label{duality}
a_2(\phi) = h_1^{\ell}(\phi), ~~ h_2(\phi) = a_1^{1/\ell}(\phi)
\ee
where $\ell$ is a constant. Cosmological models/phases related via (\ref{duality}) are said to be dual to each other. 
This property is immensely useful to establish correspondence between different cosmological phases  \cite{triality} or between 
different models \cite{mizuno,scaling}. Here we would study its symmetries for different phases of the scalar field dynamics. In the 
next section, we would apply this to establish dualities of loop cosmology with braneworld scenarios.

Let us consider an expanding cosmological phase with a standard scalar field, specified by $a_1(\phi)$ and $h_1(\phi)$. Then a simple
duality based on eq.(\ref{cllm_hub}) which is given by $h_2(\phi) = a_1(\phi)$ leads to 
\be
a_2(\phi) = h_1(\phi) = \f{\rho_1(\phi)}{\rcr - \rho_1(\phi)}
\ee
 on using eq.(\ref{aphi1}). It is then straightforward to find the rate of change of the dual scale factor,
\be
\dot a_{2}(\phi) = - \f{3 H_1 \dot \phi_1^2}{\rcr} (1 + h_1(\phi))^2 
\ee
with 
\be
H_1 =  \left(\f{\kappa \rcr}{3}\right)^{1/2} \, \f{h_1(\phi)^{1/2}}{(1 + h_1(\phi))}
\ee
and 
\be
\dot \phi_1 = - n \left(\f{\rcr}{3 \kappa} \right)^{1/2} \, \f{h_{1_{{,\phi}}}}{h_1(\phi)^{1/2}(1 + h_1(\phi))} ~.
\ee
Since, $H_1 > 0$ we obtain $\dot a_2 < 0$ and thus the duality $a(\phi) \leftrightarrow h(\phi)$ maps an expanding phase to a 
contracting phase, for the standard scalar field.

Another simple duality transformation is when $a_3(\phi) = h_1^{-1}(\phi)$ and $h_3(\phi) = a_1(\phi)$. In this case it can be checked that
$a_3(\phi)$ and $h_3(\phi)$ satisfy the Friedmann equation for the phantom field,
\be
a_3(\phi)_{{,\phi}} \, h_3(\phi)_{{,\phi}} =  \kappa \, a_3(\phi) \,  h_3(\phi) ~.
\ee
It is also easy to verify that 
\be
\dot a_3(\phi) =  \f{3 H_1 \dot \phi_1^2}{\rcr} \f{(1 + h_1(\phi))^2}{h_1(\phi)^2} 
\ee
so that if $a_1(\phi)$ corresponds to an expanding (contracting) branch then $a_3(\phi)$ also denotes an expanding (contracting) branch.
Thus, this duality transformation maps an expanding branch for the standard scalar field to an expanding branch of a phantom field.
Further, since $h(\phi)$ is greater (less) than unity for $\rho$ greater (less) than $\rcr/2$, this duality maps orbits $h_1(\phi) \gtrless 1$ 
belonging to the super-inflationary phase for the standard scalar field to the orbits $a_3(\phi) \lessgtr 1$ for the phantom field.
Similarly, the contracting branch of the standard scalar field and the phantom branch are dual to each other under the map, $a_2(\phi) = a_3^{-1}(\phi), h_2(\phi) = h_3(\phi)$.

\section{Scaling solutions and dualities with Randall-Sundrum braneworlds}
A significant issue concerning the effective dynamics in LQC is its stability. For that it is
useful to construct the scaling solutions as they provide  insights on the asymptotic behavior
of solutions and can also serve to establish symmetries between distinct cosmologies \cite{scaling}. 
In a model when only modifications due to eigenvalues of the inverse scale factor operator
are important, scaling solutions have been obtained \cite{lidsey,clm}. For the case of our interest, 
these modifications can be neglected and hence our scaling solutions would be  different from 
those in Ref. \cite{lidsey,clm}. We would consider the energy density with contributions from a massive scalar field and matter 
with a fixed equation of state $w$.
In order to study the scaling solutions it is useful to define new variables:
\bq \label{variables}
x := \left(\dot \phi^2/2\rho\right)^{1/2},  ~~ &&  
y := \left(V/\rho\right)^{1/2} 
\eq
and
\be \label{lambda}
 \Omega :=  - \frac{\kappa^{-1/2}}{(1 - \rho/\rcr)^{1/2}} \,\frac{V_{,\phi}}{V} ~.
\ee
Then the set of dynamical equations composed of eq.(\ref{mod}), the stress-energy conservation law  
\be \label{pe1}
\dot \rho_w + 3 \f{\dot a}{a} \, (\rho_w + p_w) = 0
\ee 
and the Klein-Gordon equation (\ref{kgeq1})
can be casted into dynamical equations in $(x,y,\Omega)$,
\bq
\frac{d x}{d N} &=& \label{deq1} - 3 \, x + (3/2)^{1/2} \, \Omega y^2
~ + ~ (3/2) x B(x,y,\Omega) \\
\frac{d y}{d N} &=& \label{deq2}- (3/2)^{1/2} \, \Omega x y ~ + ~
(3/2) y B(x,y,\Omega) \\
\frac{d \Omega}{d N} &=& \n \label{deq3} - \sqrt{6} x \Omega^2 V \bigg[
\frac{V_{,\phi\phi}}{V_{,\phi}^2} - 1 \bigg] \\
&& \, \, \,- \, \f{3 \Omega}{2}  B(x,y,\Omega) \f{\rho/\rcr}{(1 - \rho/\rcr)}
\eq
where $N := \log a$ and
\be
B(x,y,\Omega) := 2 x^2 + (1 + w) (1 - x^2 - y^2) ~.
\ee
As before we  also express the Friedmann equation (\ref{mod}) in the form (\ref{cllm_hub}), with $h(\phi)$ now given as
\be \label{hphi}
h(\phi) = \exp\left((x^2 + y^2) \int \f{d \rho}{\rho(1 -  \rho/\rcr)} \right) ~
\ee
where we have used $x ^2 + y^2 = \rho_\phi/\rho$. Also the scale factor $a(\phi)$ is determined by eq.(\ref{aphi1}) with the above $h(\phi)$.

The scaling solutions can then be obtained by solving
\bq
\frac{d x}{d N} &=& \n \frac{d y}{d N} = \frac{d \Omega}{d N} = 0 ~.
\eq
It is straightforward to check that for the case when $\rho_w \gg \rho_\phi$, the critical point is given by
\be
x_c = \label{critpoint1} \left(\frac{3}{2}\right)^{1/2}\, \frac{1+ w}{\Omega},  \, y_c =
\frac{1}{\Omega} \, \left(\frac{3(1- w^2)}{2}\right)^{1/2} ~
\ee
with the constraint
\be
\rho \, \frac{\rho_{,\phi\phi}}{\rho_{,\phi}^2} + \f{1}{2} \f{\rho/\rcr}{(1 - \rho/\rcr)} = \label{crit_cons} 1.
\ee
Further, when $\rho_\phi \gg \rho_w$ the critical point is
\be
x_c = \label{critpoint2} \Omega/\sqrt{6},   \, y_c = \left(1 - \f{\Omega^2}{6} \right)^{1/2} ~
\ee
with the same constraint as eq.(\ref{crit_cons}).
Interestingly, if $\Omega$ is treated as a constant then the form of the eqs.(\ref{deq1},\ref{deq2} \& \ref{deq3}) is identical to the one obtained in the standard FRW cosmology \cite{clw}. This important feature has been noted earlier
in the context of modified gravity scenarios in the string inspired cosmologies \cite{mizuno,scaling}. Due to this
correspondence with the classical equations we expect that the critical points for our case belongs to the set of critical points
in the classical theory. This turns out to be true if we recall that for classical theory, the corresponding 
variables defined as
\be
\xc = \left(\f{\kappa \dot \phi^2}{6 H^2}\right)^{1/2}, ~ \yc =  \left(\f{\kappa V}{3 H^2}\right)^{1/2}, ~ \lc = - \left(\f{V_{,\phi}}{\kappa^{1/2} V}\right)
\ee
do indeed lead to the critical points given by eq.(\ref{critpoint1}) and eq.(\ref{critpoint2}) for the cases when $\rho_w \gg \rho_\phi$ and $\rho_\phi \gg \rho_w$ respectively \cite{clw}. The stability analysis
of Ref. \cite{clw} then implies that both of the critical points (\ref{critpoint1} \& \ref{critpoint2}) are attractors if 
the former satisfies $\Omega^2 > 3 (1 + w)$ and latter satisfies $\Omega^2 < 3 (1 + w)$.

The potential for the scaling solutions can be determined by integrating eq.(\ref{crit_cons}) and using 
eq.(\ref{variables}), which turns out to be 
\be \label{lqc_pot}
V(\phi) = y_c^2 \, \rcr\,  \mathrm{sech}^2 (- \kappa^{1/2} \, \Omega \,\phi/2) ~.
\ee
On integrating eqs.(\ref{aphi1}) and (\ref{hphi}) we can obtain $a(\phi)$ and $h(\phi)$ for the two scaling solutions
\be
a(\phi) = \cosh^{2 \Omega^{-2}(x_c^2 + y_c^2)^{-1}} (-\kappa^{1/2} \Omega \phi/2)
\ee
\be
h(\phi) = \rcr^{(x_c^2 + y_c^2)} \, \mathrm{csch}^{2 (x_c^2 + y_c^2)} (-\kappa^{1/2} \Omega \phi/2) ~.
\ee

Once we have found the set $(a(\phi),h(\phi),V(\phi))$ for the scaling solutions in LQC, we can then find a model with a dual 
scaling solutions using eq.(\ref{cllm_hub}). Such dualities have been studied for modified gravity scenarios earlier and 
in Ref. \cite{scaling} it was established that if a cosmological model with a modified Friedmann equation of the form 
$H^2 = (\kappa/3) \rho(\phi) G_1(\phi)$ has a scaling solution  
$(a_1(\phi), b_1(\phi), V_1(\phi))$, then a necessary and 
sufficient condition for different cosmological model with $H^2 = (\kappa/3) \rho(\phi) G_2(\phi)$ and scaling solution 
$(a_2(\phi), b_2(\phi), V_2(\phi))$ to satisfy eq.(\ref{duality}) and to be the dual  is 
that
\be \label{dualeq}
G_1^{1/2}(\phi) = - \f{\ell f}{G_2^{1/2}(\phi)}
\ee
with 
\be 
f = (x_c^2 + y_c^2)^2  \Omega_1 \Omega_2 ~.
\ee
In order to find the dual  of the scaling
solution in LQC, we first note that $G_1(\phi) = 1 - \rho/\rcr$  which using eq.(\ref{lqc_pot}) and eq.(\ref{variables})
gives
\be
(1 - \rho/\rcr)^{1/2} = \tanh(-\kappa^{1/2} \, \Omega_1 \, \phi/2) ~.
\ee
Interestingly, the corresponding function $G_2(\phi)$  for the Randall-Sundrum scenario is \cite{scaling}
\be
(1 + \rho/2\sigma)^{1/2} = \coth(-\kappa^{1/2} \, \Omega_2 \, \phi/2)
\ee
which on using eq.(\ref{dualeq}) and further choosing $\Omega_1 = \Omega_2 = \Omega$ implies that the scaling solution of the
effective theory of LQC and the Randall-Sundrum scenario are dual to
each other if $\ell = - 1/(x_c^2 + y_c^2)^2 \Omega^2$ and $\rcr$ is
identified with the brane tension $\sigma$,
\be
\rcr = \left(3/\alpha \kappa \gamma^2 \lp^2\right) = 2 \sigma ~.
\ee
In the Randall-Sundrum braneworld scenarios $\sigma$ plays the role of a critical energy density scale near and above which
the effects due to extra dimensions become significant and the standard Friedmann dynamics is modified. In the effective 
theory of LQC, the analog of brane tension is $\rcr$ which contains the fundamental loop parameter $\gamma$. Since both 
$\rcr$ and $\sigma$ play the same role in the modified Friedmann dynamics originating from LQC and the string inspired scenarios 
respectively, the above correspondence leads to the dual relationship at the level of scaling solutions between the two scenarios.

\section{Summary}

In this work we have analyzed the effective dynamics in LQC when effects due to 
discrete quantum geometric modifications are important and those from the change in behavior of the 
eigenvalues of the inverse scale factor operator are negligible. We have shown that in this case the 
Friedmann equations gets a $\rho^2$ modification similar 
to the Randall-Sundrum scenario but with a negative sign (which also arises in a modified Randall-Sundrum model with a 
time-like extra dimension). The $\rho^2$ term is also present if the modifications  to the inverse scale factor 
for $a < a_*$ are included. 

The modification to the Friedmann equation has various interesting properties. We have shown that it leads to a generic
phase of super-inflation when $\rho > \rcr/2$. This phase is
independent of the one studied earlier \cite{superinflation,cmb,inflation} which originates due to change in behavior of the kinetic term in the Klein-Gordon equation for $a < a_*$. 
Existence of this additional phase of super-inflation in the very early Universe would increase the number of e-foldings
originating in the loop quantum modified phase and assist onset of conventional inflation. Further, an ordinary scalar field 
behaves as a phantom field and the vice versa in this regime. We have also found various symmetries linking the 
expanding and the contracting branch with the phantom field dynamics in the effective theory.

We have obtained scaling solutions in the effective theory which give us valuable information about the stability 
of the dynamics and can also be used to find symmetries between distinct cosmological models. We find that the scaling 
solutions in the effective theory are dual to those of the Randall-Sundrum scenario if the critical density 
arising in LQC is identified with the brane tension.
 Scaling
solutions for effective dynamics in LQC with arbitrary $j$ but without
any discrete quantum geometric effects have been obtained 
earlier and  dualities with standard cosmology and the string-inspired scenarios have also been 
noted \cite{lidsey,clm}. The scaling solutions obtained here and those
in Ref. \cite{lidsey,clm} thus belong to complimentary domains of the
effective theory. It would be interesting to consider effective
dynamics with discrete quantum corrections for scale factors below $a_*$
 and obtain the scaling solutions and the associated dualities with
other modified gravity scenarios. 

It is very interesting to see that two  distinct 
quantization schemes for gravity yield $\rho^2$ modification to the Friedmann equation at high energies. 
Though 
the effective theory of LQC is dual to the Randall-Sundrum braneworlds in the sense described above we shall remember 
that both models  yield different predictions for the early Universe. For example, the phase of super-inflation 
in the very early Universe and of a non-singular bounce in a contracting Universe which are the features generically 
present in LQC, are absent 
in the Randall-Sundrum scenario. Let us now point out various uses of this duality symmetry. The first and perhaps the most 
straightforward use is to apply this duality to classify the similarities and differences in various dynamical solutions 
between LQC and Randall-Sundrum braneworlds. Such a classification is important to compare physical results predicted by 
scenarios which lead to modifications of the standard Friedmann dynamics. Detailed comparison of dynamical solutions between Randall-Sundrum scenario, Dvali-Gabadadze-Porrati (DGP) braneworlds \cite{dgp} and Cardassian cosmology \cite{cardassian}
 has been done earlier \cite{scaling}. Thus with the duality established between LQC and Randall-Sundrum scenario, we can further explore the 
connection of LQC with Cardassian and DGP models, compare the dynamical properties and obtain insights into links between seemingly
unrelated and distinct cosmological models.

An interesting use of duality symmetry can be to explore the detailed dynamical properties of LQC by using the established results in braneworld scenarios. For example, 
the duality relation  can be used to calculate the spectral index of scalar perturbations in the 
effective theory in LQC. For this let us recall that recently the duality relation between inflationary and cyclic models 
was used to confirm the the value of spectral index in cyclic model, given the value in inflationary models \cite{triality}. 
In LQC, a full fledged calculation of spectral index by including inhomogeneities and reducing the symmetry of the framework 
is yet to be undertaken and the effective theory with inhomogeneities is not known. However, a calculation on the above lines starting from the spectral index in Randall-Sundrum cosmology 
and obtaining the one in LQC would provide a first estimate of what we may obtain by a 
detailed analysis. Such an exercise would also provide a test for the duality relation found in this paper as the calculation of
spectral index by using duality symmetry should confirm with the result obtained from the effective theory obtained after including 
 inhomogeneities in the quantum theory.

We should note that  in LQC, the evolution is generically non-singular and a flat  Universe bounces at scales close to Planck length \cite{aps} whereas it has remained an outstanding problem to obtain generic non-singular bouncing solutions in string inspired cosmologies.
It has been shown that duality relation can be used to relate singular cosmological background with non-singular one \cite{scaling}. This may help in shedding some light on construction of non-singular bouncing models in string inspired cosmologies by using the solutions in 
LQC. Further since LQC is based on Dirac's method of canonical quantization, another the use of this duality can be to 
establish a similar construction for braneworld scenarios.

The existence of an exact duality between the scaling solutions of LQC and the Randall-Sundrum scenario comes as a surprise since 
they are derived from two very different unrelated approaches to quantize gravity. At high energies both of the scenarios 
predict a $\rho^2$ modification to the 
Friedmann equation (although with a different sign) but the origin of this modification is the existence of an extra dimension in Randall-Sundrum scenario and the discrete 
quantum geometry in LQC. Note that the spacetime geometry in string inspired Randall-Sundrum scenario is continuous whereas LQC is based on 
a four dimensional quantization of spacetime. Existence of any symmetry relationship between such two significantly distinct models to describe 
nature of the very early Universe is very non-trivial and perhaps more than a mere coincidence. The duality symmetry between Randall-Sundrum model and LQC 
suggest that some of the effects originating due to existence of extra dimensions in a continuous spacetime bulk might be mimicked by the quantum geometric nature of a four dimensional spacetime. This signifies the non-trivial and deep nature of this duality symmetry. 
We should here emphasize that both 
string theory and LQG, the underlying theories on which Randall-Sundrum scenario and LQC are respectively based, are still far from being 
complete theories. In fact they suffer from complimentary problems: lack of a non-perturbative background independent treatment in string theory and 
little insights on the way to obtain a semi-classical perturbative description in LQG. Therefore any relation like above, if proved 
as a deep link between two theories by future investigations, might prove immensely useful 
in gaining insights on the resolution of these problems \cite{link} and 
may also provide a new paradigm for a complete theory of quantum gravity which may include both stringy and loopy ideas.

{\bf Acknowledgments:} We thank J. E. Lidsey for comments on an earlier draft of the manuscript. We also thank
A. Ashtekar, N. Dadhich,  T. Pawlowski and K. Vandersloot for discussions.
Author's work is supported in part by Eberly research funds of
Penn State and by NSF grant PHY-0456913.


\begin{thebibliography}{10}
\bibitem{wkb} Discreteness corrections have also been evaluated in the WKB approach to the effective dynamics, see \cite{shyam_discrete}.
\bibitem{formula} Formulas in LQC typically are written using $\lp = \sqrt{8 \pi G}$, hence they will appear
slightly modified when compared to previous works.
\bibitem{eff_footnote} The effective Hamiltonian is very similar to the one in Ref. \cite{eff_diff}, except that 
in the latter the value of $\bar \mu$ was not constrained by requiring that  the minimum  
physical area of the loop shall be given by the lowest eigenvalue of the area operator in LQG.
\bibitem{link} In the context of LQC, this issue has been raised earlier at different levels. See for eg. Refs. \cite{lidsey,clm,nkd}.
\end{thebibliography}
\end{document}